\documentstyle[pra,aps,twocolumn,epsf]{revtex}

\begin{document}
\draft
\title{Two-photon Diffraction and Quantum Lithography}
\author{Milena D'Angelo, Maria V. Chekhova\thanks{Permanent Address: Department
of Physics, Moscow State University, Moscow, Russia.}, and Yanhua
Shih}
\address{Department of Physics, University of Maryland, Baltimore
County, Baltimore, Maryland 21250}\maketitle

\widetext

\vspace*{-10mm}

\begin{abstract}

\noindent
We report a proof-of-principle experimental
demonstration of quantum lithography.  Utilizing the entangled
nature of a two-photon state, the experimental results have
bettered the classical diffraction limit by a factor of two. This
is a quantum mechanical two-photon phenomenon but not a violation
of the uncertainty principle.
\end{abstract}


\narrowtext
Classical optical lithography technology is facing
its limit due to the diffraction effect of light. However, this
classical limit can be surpassed, surprisingly, by utilizing the
quantum nature of entangled multi-photon states \cite{EPR}. In an
idealized experimental situation, the minimum width of the
entangled N-photon diffraction pattern can be N times narrower
than the width of a classical diffraction pattern. The working
principle of the effect has been discussed theoretically by Boto
et al \cite{Dowling}; and by Scully from a different approach
\cite{ScullyMicroscope}.  In particular, one can consider
two-photon entangled states. For a two-particle maximally
entangled EPR state, the value of an observable for either single
subsystem is indeterminate.  However, if one subsystem is
measured to be at a certain value for an observable, the value of
that observable for the other subsystem is determined with
certainty \cite{EPR}. Because of this peculiar quantum nature,
the two-photon diffraction pattern thus can be narrower than
given by the classical limit under certain conditions. This
effect has been experimentally observed by Kim and Shih
\cite{Popper}.

We wish to report a proof-of-principle quantum lithography
experiment in this letter.  By using entangled photon pairs in
Young's two-slit experiment, we found that {\em under certain
experimental conditions}, the two-photon interference-diffraction
pattern has spatial interference modulation period smaller and
diffraction pattern width narrower, by a factory of two, than in
the classical case.

One of the principles of the geometrical optics is that ``light
propagates in straight line".  If this were always true, one could
obtain the image of a physical object, for example, a physical
slit, with an unlimitedly small size by applying a perfect lens
system. Unfortunately, light is also a wave. The minimum size of
the image one can make is determined by the wave property of the
light: diffraction.  The physics is very simple: according to
Huygens's principle, each point on the primary wavefront serves as
the source of spherical secondary amplitudes (wavelets) and
advances with the speed and frequency equal to those of the
primary wave. The wavelets from a physical slit will meet at any
point in space with

\newpage \vspace*{23mm} \noindent different phases. The superposition of the
wavelets will determine the size of the image.  The intensity
distribution of light can be calculated by considering an integral
of the wavelets coming from the physical object.

\begin{figure}[hbt]
\centerline{\epsfxsize=3in
\epsffile{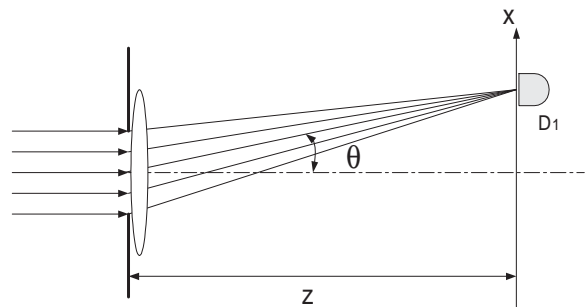}}\caption{Classical single-slit
diffraction.  Zero intensity occurs when the secondary waves
(``amplitudes") interfere destructively.}\label{Single}
\end{figure}

Figure~\ref{Single} schematically shows a classic one-dimensional
optical diffraction by a single slit. A well collimated laser beam
passes the slit and then its intensity distribution is analyzed in
the Fourier transform plane (or in the far-field zone). This
distribution, which is the diffraction pattern of a single slit,
is well-known to be ${\rm sinc}^{2}(\beta)$, where the parameter
$\beta = (\pi a/\lambda) \sin{\theta} \simeq (\pi a/\lambda)
\theta$, $a$ is the width of the slit, and $\theta$ is the angle
shown in Fig.~\ref{Single} \cite{Textbook}. When $\beta$ reaches
$\pi$, the superposition of the wavelets results in a minimum
intensity. The ${\rm sinc}^{2}(\beta)$ pattern determines the
minimum width one can obtain. Usually, this minimum width is
called the ``diffraction limit".

\begin{figure}[hbt]
\centerline{\epsfxsize=3.5in
\epsffile{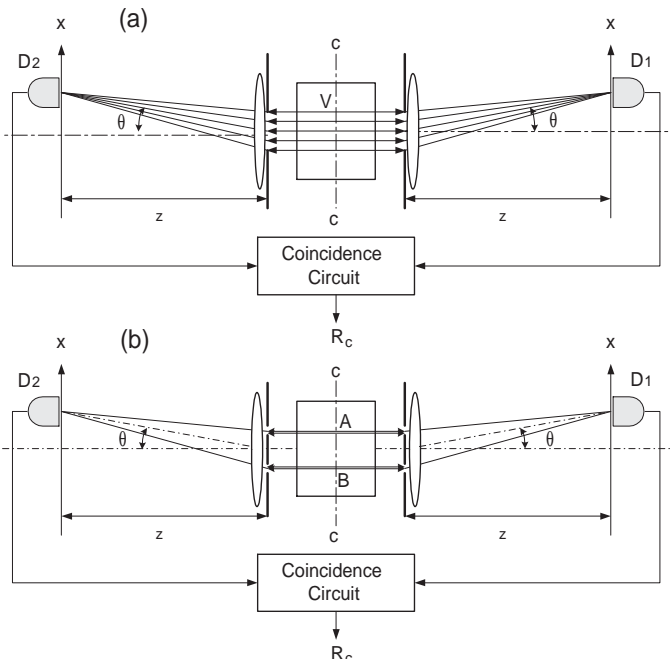}}\caption{Schematic of a two-photon
diffraction-interference {\em gedankenexperiment}. The right and
left sides of the picture represent signal and idler photons of
an entangled pair. Detectors $D_1$, $D_2$ perform the joint
detection (coincidence) measurement. It is easy to see that the
superposition of the {\em two-photon amplitudes} results in twice
faster in interference modulation and twice faster to approach
the zero intensity of diffraction than in the case of
Fig.~\ref{Single}.}\label{Scheme}
\end{figure}

To surpass the diffraction limit, our scheme is to utilize the
entangled nature of an N-particle system.  To understand the
physics of this scheme, consider a {\em gedankenexperiment} which
is illustrated in Fig.~\ref{Scheme}(a). An entangled photon pair
can be generated anywhere in region $V$; however, photons
belonging to the same pair can only propagate (1) {\em oppositely}
and (2) {\em ``almost" horizontally} (quantitative discussion will
be given later) as indicated in the figure.  Two slits are placed
symmetrically on the left and right sides of the entangled photon
source. A photon-counting detector is placed into the far-field
zone (or the Fourier transform plane, if lenses are placed
following the slits) on each side, and the coincidences between
the ``clicks" of both detectors are registered. The two detectors
are scanning symmetrically, i.e., for each coincidence
measurement, both detectors have equal x-coordinates. A two-photon
joint detection is the result of the superposition of the
two-photon amplitudes, which are indicated in the figure by
straight horizontal lines \cite{Ghost}. To calculate two-photon
diffraction, we need to ``superpose" all possible two-photon
amplitudes. Different from the classical case, it is a double
integral involving the two slits and the two-photon amplitudes
(parallel lines in Fig.~\ref{Scheme}). The two-photon counterpart
of the classical intensity, the joint detection counting rate, is
now  ${\rm sinc}^{2}(2\beta)$, which gives narrower distribution
than the classical pattern by a factor of two. Now if we ``fold"
the symmetrical left and right sides of the experimental setup
together and replace the two independent detectors with a film
that is sensitive only to two-photon light (two-photon transition
material), then in principle, we have two-photon lithography.

If one replaces the single slit in the above setup with a
double-slit, Fig. \ref{Scheme}(b), it is also interesting to see
that under the half-width diffraction pattern, the double-slit
two-photon spatial interference pattern has a higher modulation
frequency, as if the wavelength of the light were reduced to
one-half.  To observe the two-photon interference, one has to
``erase" the first-order interference by reinforcing an
experimental condition: $\delta \theta > \lambda/b$ where $\delta
\theta$ is the divergence of the light, $b$ is the distance
between the two slits, and $\lambda$ is the wavelength.

The heart of this {\em gedankenexperiment} is a special two-photon
source: the pair has to be generated in such a desired entangled
way as described above. We have found and demonstrated that, {\em
under certain conditions}, the two-photon state generated via
spontaneous parametric down conversion (SPDC) satisfies the above
requirements. The working principle, as well as the report of a
real experiment, will be given below.

\begin{figure}[hbt]
\centerline{\epsfxsize=3.25in
\epsffile{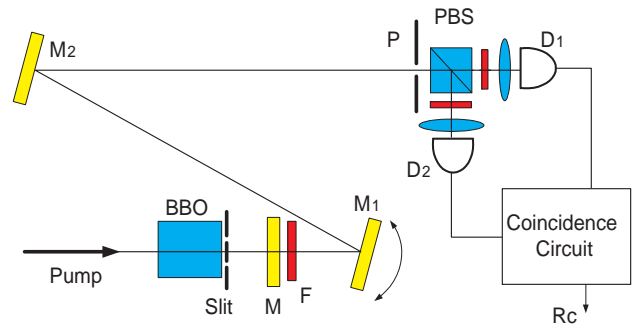}}\caption{Schematic of the experimental
setup. Entangled photon pairs are produced in a BBO crystal and
then pass through a double-slit placed immediately after the
crystal. Detectors $D_1$, $D_2$ together with the pinhole P, the
beamsplitter PBS, and the coincidence circuit represent a
two-photon detector. The angle registered by this two-photon
detector is scanned by rotating mirror $M_1$.} \label{setup}
\end{figure}

The schematic setup of the experiment is illustrated in
Fig.~\ref{setup}. It is basically the ``folded" version of a
double-slit interference-diffraction experiment shown in
Fig.~\ref{Scheme}(b).

The $458$ nm line of an Argon Ion laser is used to pump a $5mm$
BBO ($\beta-BaB_{2}O_{4}$) crystal, which is cut for degenerate
collinear type-II phase matching \cite{Klyshko,Yariv,type} to
produce pairs of orthogonally polarized signal (e-ray of the BBO)
and idler (o-ray of the BBO) photons. Each pair emerges from the
crystal collinearly, with $\omega _{s}\simeq \omega _{i}\simeq
\omega _{p}/2,$ where $\omega _{j}$ ($j=s,i,p$) are the
frequencies of the signal, idler, and pump, respectively. The
pump is then separated from the signal-idler pair by a mirror M,
which is coated with reflectivity $R\simeq 1$ for the pump and
transmissivity $T\simeq 1$ for the signal-idler. For further pump
suppression, a cutoff filter F is used. The signal-idler beam
passes through a double-slit~\cite{double}, which is placed close
to the output side of the crystal, and is reflected by two
mirrors, $M_{1}$ and $M_2$, onto a pinhole P followed by a
polarization beam splitter PBS. The signal and idler photons are
separated by the beam splitter and are detected by the photon
counting detectors $D_1$ and $D_2$, respectively. The output
pulses of each detector are sent to a coincidence counting
circuit with a $1.8ns$ acceptance time window for the
signal-idler joint detection. Both detectors are preceded by
$10nm$ bandwidth spectral filters centered at the degenerate
wavelength, $916nm$. The whole block containing the pinhole, PBS,
the detectors, and the coincidence circuit can be considered as a
two-photon detector. Instead of moving two detectors together as
indicated in Fig.~\ref{Scheme}, we rotate the mirror $M_{1}$ to
``scan" the spatial interference-diffraction pattern relative to
the detectors.

One important point to be emphasized is that the double-slit must
be placed {\em as close as possible} to the output surface of the
BBO crystal. Only in this case, can the observed diffraction
pattern be narrower than in the classical case by a factor of 2,
see Eq.~(\ref{condition2}). Otherwise, it will be close to
$\surd2$ as suggested in Ref.~\cite{ScullyMicroscope}.

\begin{figure}[hbt]
\centerline{\epsfxsize=3.5in
\epsffile{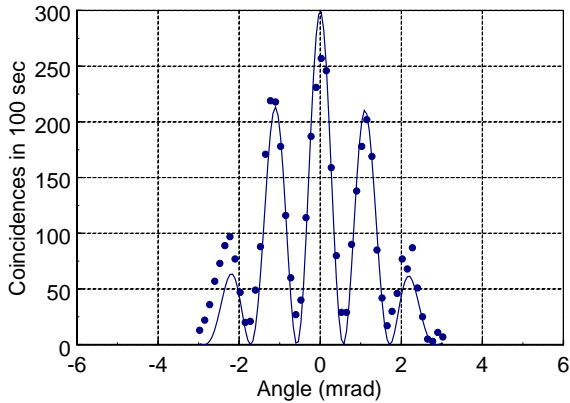}}\caption{Experimental measurement of a
two-photon double-slit interference-diffraction pattern.
Comparing with the classical case, it has a faster spatial
interference modulation and a narrower diffraction pattern width,
by a factor of 2. }\label{Doubledata}
\end{figure}

Figure \ref{Doubledata} reports the experimental results.  In our
experiment, the width of each slit is $a = 0.13mm$. The distance
between the two slits is $b = 0.4mm$.  The distance between the
slits and the pinhole P is $4m$. Figure \ref{Doubledata} shows the
distribution of coincidences versus the rotation angle $\theta$ of
mirror $M_1$. The spatial interference period and the first zero
of the envelope are measured to be $0.001$ and $\pm0.003$ radians,
respectively.

\vspace*{-10mm}

\vspace*{-10mm}
\begin{figure}[hbt]
\centerline{\epsfxsize=3.5in
\epsffile{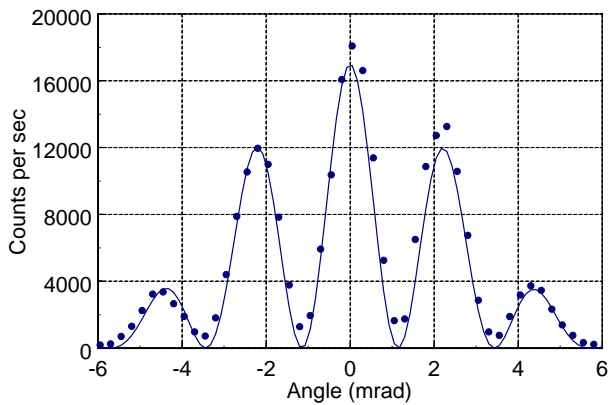}}\caption{Measurement of the
interference-diffraction pattern for classical light in the same
experimental setup.}\label{Classicaldata}
\end{figure}

For comparison, we also measured the first-order
interference-diffraction pattern of a classical light with $916nm$
wavelength by the same double-slit in the same experimental setup,
see Fig.~\ref{Classicaldata}.  The spatial interference period and
the first zero of the envelope are measured to be $0.002$ and
$\pm0.006$ radians, respectively.

In both ``classical" and ``quantum" cases, we obtain similar
standard Young's two-slit interference-diffraction pattern,
${\rm sinc}^2[(\pi a/\lambda)\theta] \cos^2[(\pi
b/\lambda)\theta]$; however, whereas the wavelength for fitting
the curve in Fig.~\ref{Classicaldata} (classical light) is
$916nm$, for the curve in Fig.~\ref{Doubledata} (entangled
two-photon source) it has to be $458nm$.  Clearly, the two-photon
diffraction ``beats" the classical limit by a factor of 2.

To be sure that we observed the effect of the SPDC photon pair
with wavelength of $916nm$ but not the pump laser beam with
wavelength of $458nm$, we remove or rotate the BBO crystal
$90$-degree to a non-phase-matching angle and examine the
coincidence counting rate.  The coincidences remain zero during
the $100$ second period, which is the data collection time
duration for each of the data points, even in high power operation
of the pump laser. Comparing this with the coincidence counting
rate with BBO under phase-matching, see Fig.~\ref{Doubledata},
there is no doubt that the observation is the effect due to the
SPDC.

To explain the result, we have to take into account the quantum nature of the
two-photon state. SPDC is a nonlinear optical process in which pairs of
signal-idler photons are generated when a pump laser beam is incident onto an
optical nonlinear material \cite{Klyshko,Yariv}. Quantum mechanically, the
state can be calculated by the first-order perturbation theory \cite{Klyshko}
and has the form
\begin{eqnarray}\label{state1}
\left| \Psi \right\rangle =\sum_{s,i} F(\omega _s,\omega _i,{\bf k}_s,{\bf k}_i)
a_{s}^{\dagger }(\omega ({\bf k}_{s}))\ a_{i}^{\dagger }(\omega
({\bf k}_{i}))\mid 0\rangle,
\end{eqnarray}
where $\omega _{j}$, {\bf k$_{j}$ (}j = s, i, p) are the
frequencies and wavevectors of the signal (s), idler (i), and pump
(p), respectively, $F(\omega _s,\omega _i,{\bf k}_s,{\bf k}_i)$ is
the so-called biphoton amplitude, the pump frequency and
wavevector $\omega _{p}$ and {\bf k}$_{p}$ can be considered as
constants, $a_{s}^{\dagger }$ and $a_{i}^{\dagger }$ are creation
operators for the signal and idler photons, respectively. The
biphoton amplitude contains $\delta$-functions of the frequency
and wavevector,
\begin{equation}
\label{delta} F(\omega _s,\omega _i,{\bf k}_s,{\bf k}_i)\propto \delta \left(
\omega _{s}+\omega _{i}-\omega _{p}\right) \delta \left( {\bf k}_{s}+{\bf
k}_{i}-{\bf k} _{p}\right).
\end{equation}
The signal or idler photon could be in any modes of the
superposition (uncertain); however, due to Eq.~(\ref{delta}), if
one photon is known to be in a certain mode then the other one is
determined with certainty.

The phase matching conditions resulting from the $\delta $-functions in
Eq.~(\ref{delta}),
\begin{equation}  \label{phasematch}
\omega _s+\omega _i=\omega _p,\quad {\bf k}_s{\bf +k}_i={\bf k}_p,
\end{equation}
play important role in the experiment.  The transverse component
of the wavevector phase matching condition requires that
\begin{equation}  \label{anglein}
k_s\sin \alpha _s =k_i\sin \alpha _i,
\end{equation}
where $\alpha _s$ and $\alpha _i$ are the scattering angles inside the crystal.
Upon exiting the crystal, Snell's law thus provides:
\begin{equation}
\omega _{s}\sin \beta _{s}=\omega _{i}\sin \beta _{i}
\label{angleout}
\end{equation}
where $\beta _{s}$ and $\beta _{i}$ are the exit angles of the
signal and idler with respect to ${\bf k}_{p}$ direction.
Therefore, in the degenerate case, the signal and idler photons
are emitted at equal, yet opposite, angles relative to the pump,
and the measurement of the momentum (wavevector) of the signal
photon determines the momentum (wavevector) of the idler photon
with unit probability and vice versa.  In the collinear case,
which we use in our experiment, the scattering angles of the
signal and idler photons are  close to zero and occupy the range
$\Delta\theta$, which is determined by the sizes of the crystal
and the pump beam, see \cite{BigPaper}.

The coincidence counting rate $R_c$ is given by the probability $P_{12}$ of
detecting the signal-idler pair by detectors $D_1$ and $D_2$ jointly,
\begin{eqnarray}  \label{P12}
P_{12}&=&\left\langle \Psi \right| \
E_1^{(-)}E_2^{(-)}E_2^{(+)}E_1^{(+)}\ \left| \Psi \right\rangle
\\ \nonumber &=&\left| \left\langle 0\right| \ E_2^{(+)}E_1^{(+)}\
\left| \Psi \right\rangle \right| ^2,
\end{eqnarray}
where $\left| \Psi \right\rangle $ is the two-photon state of SPDC and $E_1$,
$E_2$ are fields on the detectors. The effect of two-photon Young's
interference can be easily understood if we assume for simplicity that signal
and idler photons always go through the same slit and never go through
different slits. This approximation holds if  the variation of the scattering
angle inside the crystal
\begin{equation}\label{condition}
\Delta \theta \ll b/D,
\end{equation}
where $D$ is the distance between the input surface of the SPDC crystal and the
double-slit. In this case, the state after the slits can be written as
\begin{equation}  \label{state2}
\left| \Psi \right\rangle =\left| 0\right\rangle +\epsilon \
[a_s^{\dagger }a_i^{\dagger }\exp (i\varphi _A)\ +\ b_s^{\dagger
}b_i^{\dagger }\exp (i\varphi _B) \left]| 0\right\rangle,
\end{equation}
where $\epsilon \ll 1$ is proportional to the pump field and the
nonlinearity of the crystal, $\varphi _A$ and $\varphi _B$ are the
phases of the pump field at region A (upper slit) and region B
(lower slit), and $a_j^{\dagger }$ ($b_j^{\dagger }$) are the
photon creation operators for photons passing through the upper
slit (A) and the lower slit (B). In our experiment, the ratio
$(b/D)/\Delta \theta \simeq 30$ and Eq.~(\ref{condition}) is
satisfied well enough. Moreover, even the ratio
$(a/D)/\Delta\theta$ is of order of 10, which satisfies the
condition for observing two-photon diffraction:
\begin{equation}\label{condition2}
\Delta \theta \ll a/D.
\end{equation}

In Eq.~(\ref{P12}), the fields on the detectors are given by:
\begin{equation}  \label{field}
\begin{array}{c}
E_1^{(+)}=a_s\exp (ik\ r_{A1})\ +\ b_s\exp (ik\ r_{B1}) \\
E_2^{(+)}=a_i\exp
(ik\ r_{A2})\ +\ b_i\exp (ik\ r_{B2})
\end{array}
\end{equation}
where $r_{Ai}$ ($r_{Bi})$ are the optical path lengths from region A (B) to the
ith detector. Substituting Eqs.~(\ref {state2}) and (\ref{field}) into
Eq.~(\ref{P12}), we get
\begin{eqnarray}
R_c \propto P_{12}&=& \epsilon ^2\left| \exp (ik\ r_A+i\varphi
_A)\
+\exp (ik\ r_B+i\varphi _B)\ \right| ^2 \nonumber\\
&\propto& 1+\cos [k\ (r_A-r_B)],\label{Rc1}
\end{eqnarray}
where we define $r_A \equiv r_{A1}+r_{A2}$. We have assumed $\varphi_A=\varphi
_B$ in Eq.~(\ref{Rc1}).

In the far-field zone (or the Fourier transform plane), interference of the two
amplitudes from Eq.~(\ref {state2}) gives
\begin{equation}  \label{Rc2}
R_c(\theta)\propto \cos {}^2[(2\pi b/\lambda)\theta].
\end{equation}
Eq.~(\ref{Rc2}) has the form of a standard Young's two-slit
interference pattern, except having the modulation period one-half
of the classical case or an equivalent wavelength of $\lambda/2$.

To calculate the diffraction effect of a single slit, we need an integral of
the effective two-photon wavefunction over the slit width. Quite similarly to
Eq.~(\ref{Rc2}), it gives
\begin{equation}  \label{RcD1}
R_c(\theta)\propto {\rm sinc}^2[(2\pi a/\lambda)\theta].
\end{equation}
Eq.~(\ref{RcD1}) has the form of standard single-slit diffraction pattern,
except having half of the classical pattern width.

The combined interference-diffraction coincidence counting rate for the
double-slit case is given by
\begin{equation}  \label{RcD2}
R_c(x)\propto {\rm sinc}^2[(2\pi a/\lambda)\theta) \cos {}^2[(2\pi
d/\lambda)\theta],
\end{equation}
which is a product of Eq.~(\ref{Rc2}) and Eq.~(\ref{RcD1}).

The experimental observations have confirmed the above quantum
mechanical predictions.

In conclusion, we have demonstrated the possibility of quantum
lithography by using an entangled two-photon state generated from
a specially designed spontaneous parametric down conversion. One
may not see advantages from the above proof-of-principle
experimental demonstration. The results from the two-photon
diffraction measurement is equivalent to that of using the pump
laser beam directly.  The advantage, however, is in the case of
large number of entangled particle states.  We have considered
quantum lithography based on our entangled N-photon scheme
($N\geq3$)\cite{3-photon}. In these cases one can ``beat" the
classical limit by a factor of $N$ and keep the ``pump" laser beam
wavelength close to one-half that of the entangled photon beam.

The authors thank Y.H. Kim, J.P. Dowling, D.V. Strekalov, M.O.
Scully and H.S. Pilloff for helpful discussions.  This work was
supported, in part, by ARDA-NSA, ONR, and NSF. M.V. Chekhova
acknowledges the support from the Russian Foundation for Basic
Research, grant No. 99-02-16419.


\begin{references}
\bibitem{type}  In type-I SPDC, signal and idler are both ordinary (or
extraordinary) rays of the crystal; however, in type-II SPDC they are
orthogonally polarized, i.e., one is ordinary and the other is
extraordinary.
\bibitem{double} We use a double-slit instead of a single
slit since the interference-diffraction pattern has more details in this case
and is therefore more ``recognizable".
\end{references}
\end{document}